\documentclass[journal]{IEEEtran}
\usepackage{color}
\usepackage{graphicx}
\usepackage{subfigure}
\usepackage{verbatim}
\usepackage{url}

\ifCLASSINFOpdf
\else
\fi

\usepackage{amsmath}

\begin{document}
\title{Measuring Intra-pixel Sensitivity Variations of a CMOS Image Sensor }

\author{Swaraj Bandhu Mahato,~\IEEEmembership{Member,~IEEE,}
        Joris De Ridder, Guy Meynants, Gert Raskin,
        and~Hans Van Winckel% <-this % stops a space
\thanks{S. B. Mahato, J. De Ridder, G. Raskin,and~H. Van Winckel are at the Instituut voor Sterrenkunde, KU Leuven, Celestijnenlaan 200D, B-3001 Leuven, Belgium (e-mail: swarajbandhu.mahato@kuleuven.be).}% <-this % stops a space
\thanks{G. Meynants is at AMS, Coveliersstraat 15, B-2600 Antwerpen, Belgium.}% <-this % stops a space
\thanks{Manuscript received Month Date, 2017; revised Month Date, 2017.}}

% The paper headers
\markboth{IEEE SENSORS JOURNAL,~Vol.~xx, No.~xx, Month~2017}%
{Shell \MakeLowercase{\textit{et al.}}: Measuring Intra-pixel Sensitivity Variations of a CMOS Image Sensor}

% make the title area
\maketitle
\begin{abstract}
Some applications in scientific imaging, like space-based high-precision photometry, benefit from a detailed characterization of the sensitivity variation within a pixel. A detailed map of the intra-pixel sensitivity (IPS) allows to increase the photometric accuracy by correcting for the impact of the tiny sub-pixel movements of the image sensor during integration. This paper reports the measurement of the sub-pixel sensitivity variation and the extraction of the IPS map of a front-side illuminated CMOS image sensor with a pixel pitch of 6 \boldmath$\mu m$. Our optical measurement setup focuses a collimated beam onto the imaging surface with a microscope objective. The spot was scanned in a raster over a single pixel to probe the pixel response at each (sub-pixel) scan position. We model the optical setup in ZEMAX to cross-validate the optical spot profile  described by an Airy diffraction pattern. In this work we introduce a forward modeling technique to derive the variation of the IPS. We model the optical spot scanning system and discretize the CMOS pixel response. Fitting this model to the measured data allows us to quantify the spatial sensitivity variation within a single pixel. Finally, we compare our results to those obtained from the more commonly used Wiener deconvolution.
\end{abstract}

% Note that keywords are not normally used for peerreview papers.
\begin{IEEEkeywords}
CMOS image sensor, Detectors, Sensitivity, Intra-pixel, Astronomical instrumentation.
\end{IEEEkeywords}

% For peer review papers, you can put extra information on the cover
% page as needed:
% \ifCLASSOPTIONpeerreview
% \begin{center} \bfseries EDICS Category: 3-BBND \end{center}
% \fi
%
% For peerreview papers, this IEEEtran command inserts a page break and
% creates the second title. It will be ignored for other modes.
\IEEEpeerreviewmaketitle

%---------------------
\section{Introduction}
%---------------------

\IEEEPARstart{T}{he} spatial variation of the sensitivity within a pixel of a solid state imager can affect the total flux measured by the imager. The knowledge of this IPS map or pixel response function (PRF) \cite{IPSVref:R1}, \cite{IPSVref:R20}. \cite{IPSVref:R21} is of great interest for scientific imaging in astronomy. For example, space missions designed to detect exo-planets and stellar oscillations monitor the same star field for a long period of time, continuously making high-precision photometric measurements of each star in the field. Due to the spacecraft pointing jittering, the star image continuously moves over the sensor causing extra photometric noise because of IPS variations \cite{IPSVref:R2}. This spatial variation in sensitivity to photons depends on the manufacturing process, depth of the depletion layer and intra-pixel quantum efficiency variation. Additionally, because of the complexity of the pixel structure of the solid-state detector, numerous effects resulting from reflection and refraction at the interfaces between the layers and microlens imperfection can cause its response to vary significantly over the area of a single pixel. IPS characterization of the astronomical detectors has undergone a continuous study during the past decade. For example, In \cite{IPSVref:R3}, intra-pixel response of an infrared detector for the James Webb Space Telescope (JWST) is estimated . In \cite{IPSVref:R2}, Bryson et al. reported a study of the Kepler pixel response function. \cite{IPSVref:R4} presents a new Continuous Self-Imaging Grating (CSIG) technique to measure the IPS map of the CCD in the framework of Euclid mission. Most prior publications studied extensively the IPS variation profiles of front and back illuminated CCD’s \cite{IPSVref:R5,IPSVref:R6,IPSVref:R7,IPSVref:R8}, but so far, few studies have been done on CMOS image sensors \cite{IPSVref:R24}. 

To understand the high-precision scientific imaging capability of the latter, one must therefore characterize their IPS variation. In this paper we introduce a forward modeling technique to measure the sub-pixel sensitivity variation of a front-side illuminated CMOS image sensor. Many of the previous works in the literature derive the IPS map by performing a deconvolution (backward modeling) in which the measurement uncertainty is not taken into account. A forward modeling approach can consider the measurement uncertainty \cite{IPSVref:R9,IPSVref:R10}. In this work we model the spot projection and discretize the pixel into intra-pixels. We fit the measured data to the model of the pixel image, by minimizing the goodness-of-fit (chi-square) using a modified version of the non-linear Levenberg-Marquardt algorithm \cite{IPSVref:R11}, and estimate the sensitivity of each of the intra-pixels.

%---------------------------
\section{EXPERIMENT DETAILS}
%---------------------------

\subsection{Experimental Setup}
\begin{figure}[ht]
\centering
\includegraphics[width=0.65\linewidth]{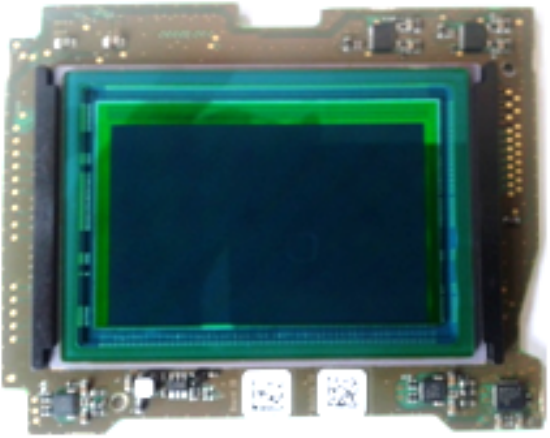}
\caption{$7920\times5136$ pixel CMOS image sensor with 45x30 $mm^2$ image area. Pixel size is $6 \mu m$ square.}
\label{fig:CMOS_sensor}
\end{figure}

\begin{figure}[ht]
\centering
\includegraphics[width=0.5\linewidth]{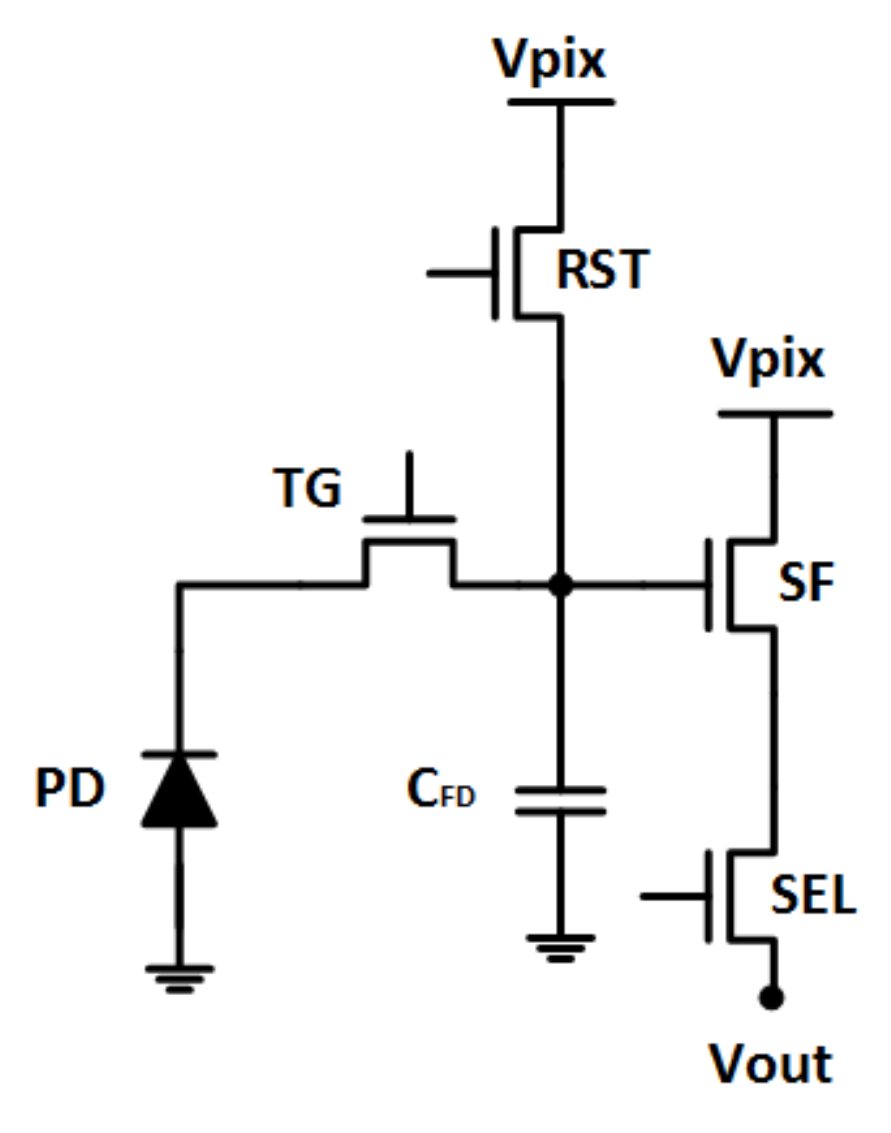}
\caption{Schematic diagram of a 4T CMOS pixel.}
\label{fig:4Tpixel}
\end{figure}

%------------------------------
We investigate the intra-pixel sensitivity variation of a front-illuminated CMOS color image sensor (Fig. \ref{fig:CMOS_sensor}) developed by CMOSIS. This detector has $7920 \times 5136$ pixels, each 6 $\mu m$ square. The state-of-the-art 4T (four-transistors) pixel design includes the correlated double sampling.   
Fig. \ref{fig:4Tpixel} shows the schematic diagram of the pixel. Each 4T pixel consists of a pinned photodiode (PD), a reset transistor (RST), a transfer gate (TG), a source follower (SF) amplifier, a row selection transistor (SEL) and floating diffusion parasitic capacitor ($C_{FD}$).

\begin{figure}[ht]
\centering
\includegraphics[width=0.9\linewidth]{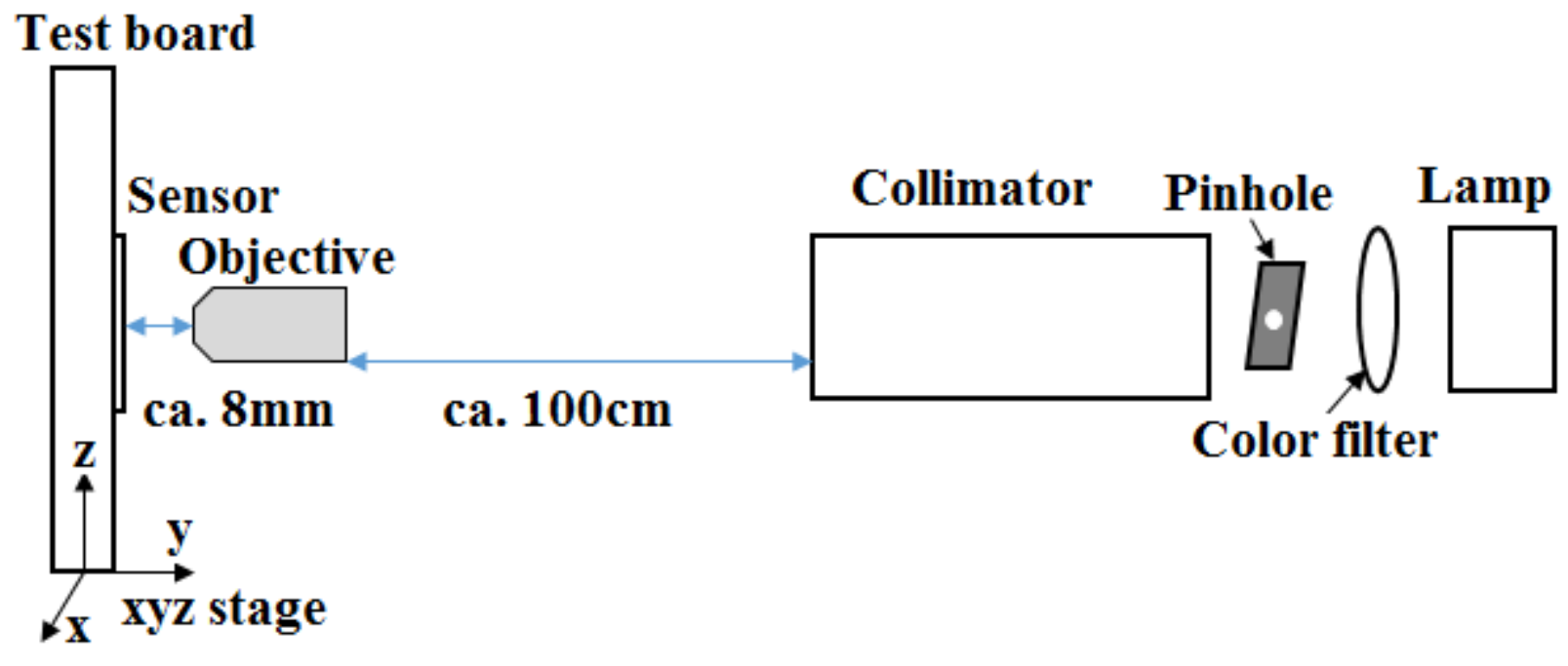}
\caption{Schematic view of the optical setup.}
\label{fig:exp_setup}
\end{figure}

We used an optical setup to project a small circular spot. The setup focuses a collimated beam onto the image sensor using a reverse telescope and a Carl Zeiss GF-Planachromat (12.5x) microscope objective. A more detailed description of the optical setup is given in \cite{IPSVref:R9}. All measurements were performed with a pinhole of $20 \mu m$. The light source is a $250 W$ incandescent lamp, mounted inside a housing that contains a lens at its exit port. The lamp is positioned such that the image of the glowing filament is in focus at the object. The spot was scanned in a raster over a single pixel to probe the pixel response at each (sub-pixel) scan position. The image sensor was mounted on a high-precision compact linear stage. We used frame grabber software on a PC and a cameralink interface to capture the experimental data. Fig.~\ref{fig:exp_setup} shows a schematic view of the optical measurement setup.

The system we used is diffraction limited. Because of the circular shape of the microscope objective in our measurements, we consider the beam profile as an Airy diffraction pattern \cite{IPSVref:R12}, which defines the instrumental point spread function (iPSF). The Abbe limit of the microscope objective is given by:
\begin{equation}
r=\frac{1.22 \lambda}{2.NA}
\label{eq:1}
\end{equation}
Where $r$ is the radius of the spot (distance from center to first minimum of the Airy disk), $\lambda$ is the wavelength and $NA=0.25$ is the numerical aperture. The calculated radius of the spot of our setup is therefore $1.34 ~\mu m$ using a green filter with $\lambda= 550 ~nm$.

\subsection{Experimental Procedure}
%----------------------------------
The experimental goal is to scan a target pixel by the projected optical spot to generate pixel response data for each sub-pixel position. Before we start the scan we need to position and focus the optical spot to the center of the pixel of interest (POI) and then move to the starting position (upper left corner) of the scan area. The center of the POI is obtained by finding the spot position where neighboring pixels get the equal amount of pixel counts (light). After that the Y-axis of the motion controller (see Fig. \ref{fig:exp_setup}) was adjusted to set the detector in the focus plane of the microscope objective when all neighboring pixels get equal and the least amount of light. Both center positioning and focusing is done with multiple iterations when images were grabbed continuously. 

\begin{figure}[ht]
\centering
\includegraphics[width=0.8\linewidth]{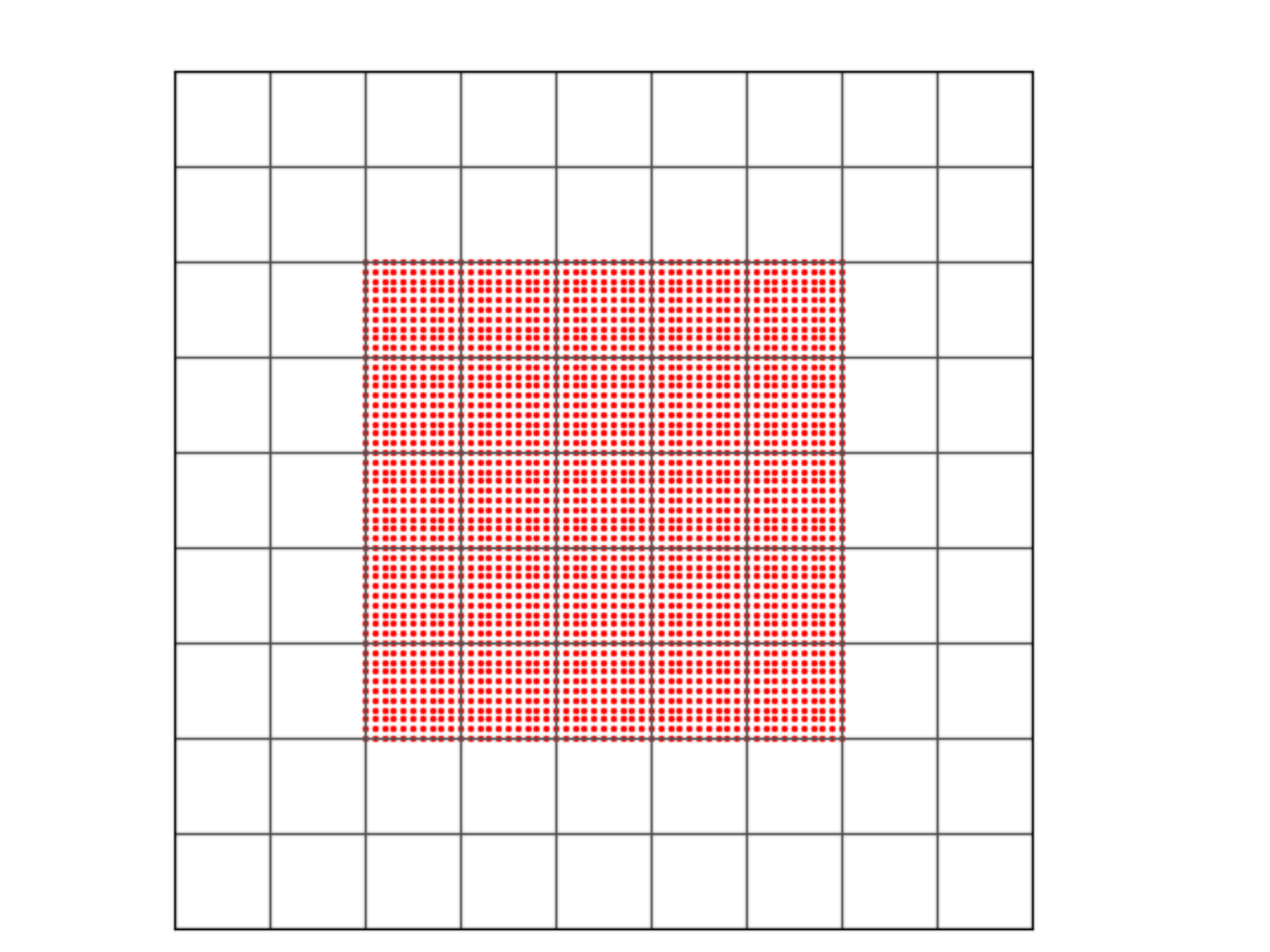}
\caption{The scanning geometry of the experiment. The sensor was scanned over a $5\times 5$ pixel area (red dots are the $51\times 51$ scan positions) and the captured sub-frame images for each spot position have a $9\times 9$ pixel area.}
\label{fig:ScanGeo}
\end{figure}

The sensor was scanned over a $5\times 5$ pixel area where POI is the center pixel. After finding the center, we move the optical spot $15~\mu m$ left and $15~\mu m$ up to find the upper left corner of the scan area as the the starting position of the scan. To avoid any thermal drift of the optical spot in X, Y or Z direction, we pre-illuminated the lamp for 4 hr before starting the scan. The optical spot is relocated across the detector and multiple sub-frame images for each spot position are captured.  We selected $9\times 9$ pixel area around the POI for which data is captured. The scanning geometry is depicted in Fig. \ref{fig:ScanGeo}. The pixel size of the detector is $6 ~\mu m \times 6 ~\mu m$ and the scanning step-size is selected as $0.6 ~\mu m$. So, a single scan line is comprised of 10 steps per pixel for a total of ($50+1$) scan points, and this for 5 pixels. At each scan point we carefully selected the exposure time to keep a high SNR (Signal to Noise Ratio) while avoiding saturation ($<16000$ ADU (Analog to Digital Units)) and nonlinearity.

For each scan point, we captured 12 sub-frames and calculated the averaged frame to reduce the noise. The scanning area contains $2601 (= 51 \times 51)$ measurements in total. All analysis was done using the Python scripting language.

%-----------------------
\section{The spot model}
%-----------------------
The beam spot profile is described in the form of Airy diffraction pattern. The intensity of the instrumental PSF at $(x,y)$ depends on the distance from the chosen spot position $(x_k, y_k)$ and on the first order Bessel function of the first kind:
\begin{equation}
I(x,y | I_0, x_k, y_k) = I_0 \left[\frac{2 J_1(\frac{\pi r}{R/R_z})}{\frac{\pi r}{R/R_z}}\right]^2
\label{eq:2}
\end{equation}
where $I_0$ refers to the peak pixel value, $J_1$ is the first order Bessel function of the first kind, $r = \sqrt{(x - x_k)^2 + (y - y_k)^2}$ is the radial distance from the maximum of the Airy disk, $R$ is the input radius parameter which is calculated as $R = 1.34 ~\mu m$ at a wavelength of $550 ~nm$ for the used microscope objectives ($NA=0.25$), and $R_z = 1.219$, is the first zero of $J_1\pi^{-1}$. For an optical system, the radius of the first zero represents the limiting angular resolution. 

To gain confidence whether the instrumental point spread function can indeed be described by Eq.~\ref{eq:2}, we model the  optical setup in ZEMAX \cite{IPSVref:R23}.  As the collimator optics of the measurement setup is unknown, we used a disk shaped light source to generate a parallel beam. We choose a non-sequential \textit{source ellipse}, which acts as an elliptical surface that emits light from a virtual point source. The parameters of the source ellipses are calculated using the sequential field locations and entrance pupil diameter. The parallel beam is then focused by an aspheric lens. The lens surface has the form of a conic section rotated about the optical axis, defined by the radius $R$ and the conic section parameter $k$. The combination of the parallel beam and the aspheric lens is a good description of the collimated beam and the microscope objective of the measurement setup. Using the ZEMAX Optimization Tool, we adjust $R$ and $k$ such that the RMS radius of the focused beam spot is minimized and that the focal distance is equal to the working distance of the microscope objective. 

\begin{figure}[ht] 
\centering \includegraphics[width=0.9\linewidth]{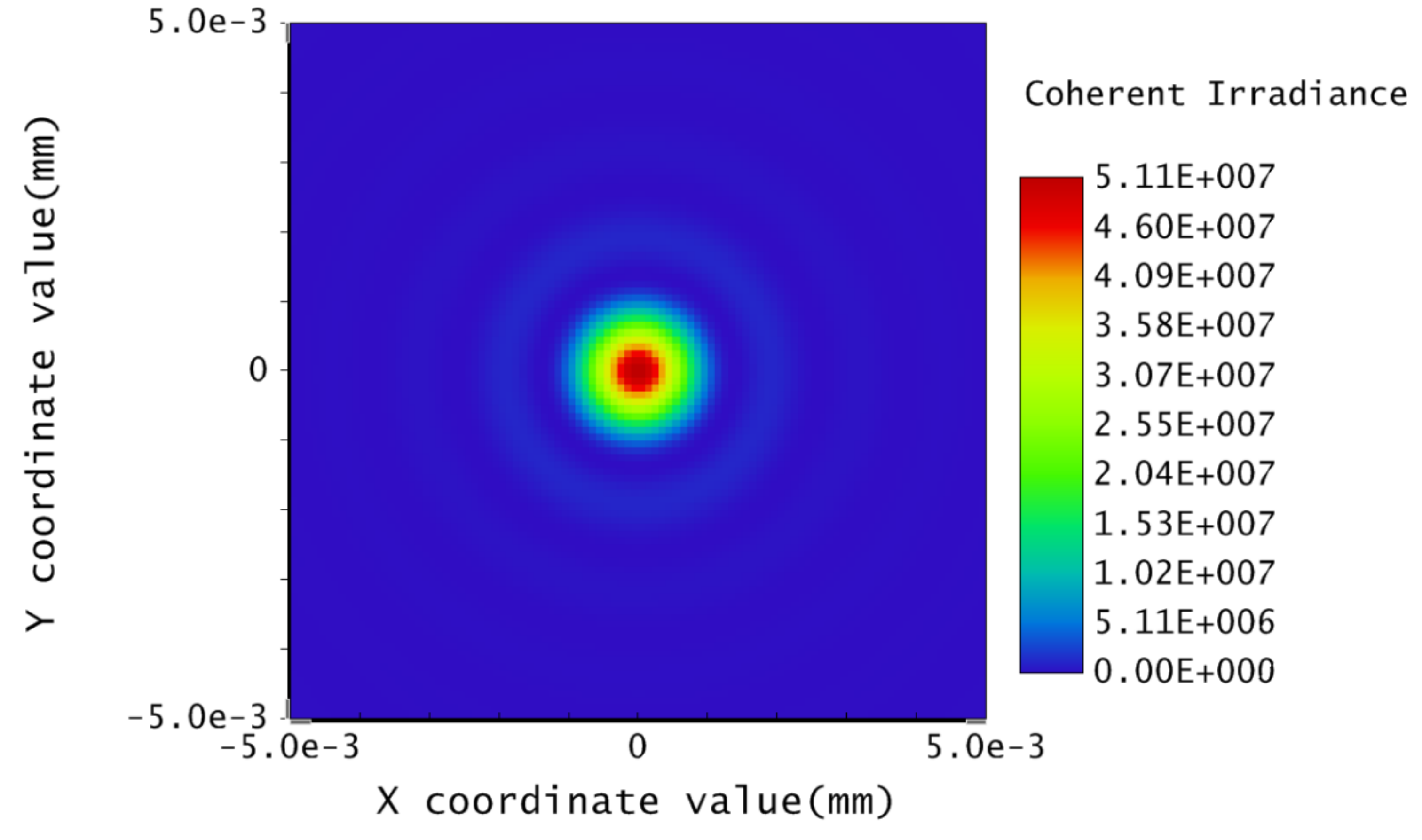}
\caption{\label{fig:psfNSC} Simulated 2D spot profile using a ZEMAX model.}
\end{figure}

\begin{figure}[ht] 
\centering \includegraphics[width=0.9\linewidth]{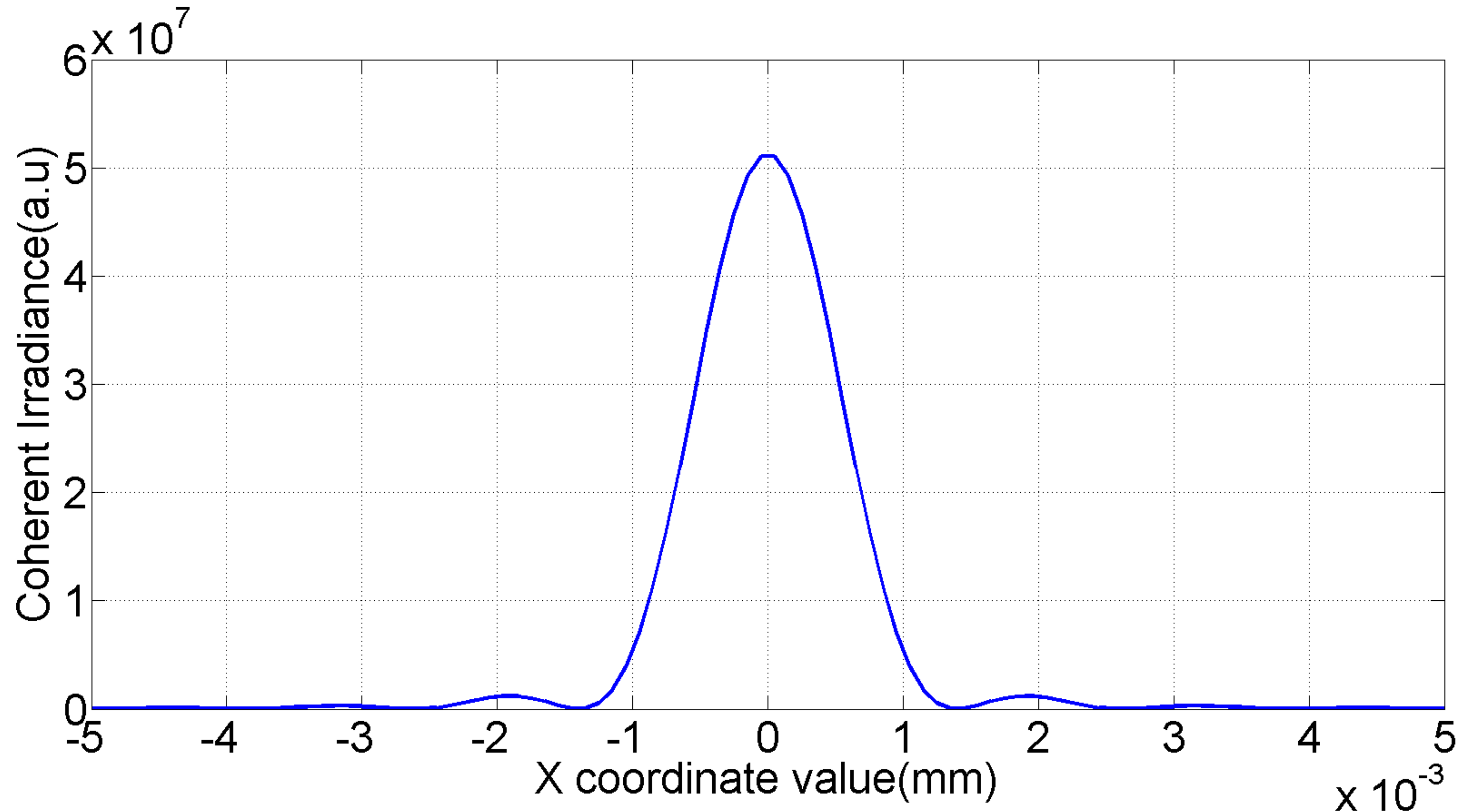}
\caption{\label{fig:1DpsfNSC} Linear profile of of the simulated spot.}
\end{figure}

Given the optimized parameters, Fig.~\ref{fig:psfNSC} shows the intensity distribution in the focus plane, perpendicular to the optical axis. 
Fig.~\ref{fig:1DpsfNSC} presents the linear profile of this simulated spot through the maximum. It shows that the radius of the spot (distance from center to the first minimum) in the simulated profile is around $1.35~\mu m$ which is in agreement with the value calculated from Eq.~(\ref{eq:1}).
Our ZEMAX simulation therefore confirms that the optical spot can be represented by an Airy disk with $1.34 ~\mu m$ radius.

%---------------------------------
\section{The Pixel Response Model}
%---------------------------------
Our CMOS image sensor consists of a 2D rectangular array of a few million pixels, each having an area of $6~\mu m$ square. To model the intra-pixel sensitivity variations, we discretize each pixel into $r \times r$ square sub-pixels, where we determine the optimal value for $r$ in a statistical sense. Each sub-pixel $(i,j)$ with $i,j=1,\cdots,r$, has its own sensitivity $S_{ij}$. The expected pixel response of the POI for a given spot position $(x_k, y_k)$ is then given by
\begin{eqnarray}
M(x_k,y_k;I_0;\{S_{ij}\}) \nonumber\\
=\sum_{i,j}^{r\times r}S_{ij}\int_{y_j^{\rm min}}^{y_j^{\rm max}}\int_{x_i^{\rm min}}^{x_i^{\rm max}}I(x,y | I_0,x_k,y_k) dx dy
\label{eq:3}
\end{eqnarray}
where the sum runs over all sub-pixels, $y_j^{\rm min}$, $y_j^{\rm max}$, $x_i^{\rm min}$, and $x_i^{\rm max}$ define the boundaries of the sub-pixel $(i,j)$, and where $I(x,y | I_0,x_k,y_k)$ is the Airy disk function given its peak value $I_0$ and its location $(x_k, y_k)$ as defined in Eq.~(\ref{eq:2}). 

\subsection{IPS Map Fitting Procedure}
%-------------------------------------
We extract the IPS map $\{S_{ij}\}$ by fitting the observed pixel responses with the expected ones given by Eq.~(\ref{eq:3}), using a weighted $\chi^2$ goodness-of-fit measure:
\begin{equation}
\chi^2 = \sum_{k=1}^{N}\left[\frac{I_{{\rm obs},k} - M_k(x_k,y_k;I_0;S)}{\frac{\sigma_k}{\sqrt{12}}}\right]^2
\label{eq:Chi}
\end{equation}
Here, $N$ is the total number of spot positions. The observed/measured data, $I_{{\rm obs},k}$ is the pixel response of the POI, captured for $k^{th}$ spot position. $\sigma_k$ is the uncertainty of the measurement. As the observed image is the average of 12 frames, the uncertainty of the measured data $I_{{\rm obs},k}$ is lowered by factor of $\sqrt{12}$. The $r\times r$ sensitivity parameters $S_{ij}$ (the optimal value for $r$ is determined in the next Section) and the Airy peak intensity $I_0$ are the fit parameters. For the least-squares fitting procedure we used a modified version of the non-linear Levenberg-Marquardt algorithm \cite{IPSVref:R11}. Obviously, the predicted pixel responses of our best-fit model (\ref{eq:3}) should match closely the measured values. We select $\sigma_k$, the uncertainty of the measured $I_{{\rm obs},k}$ (in digital number, DN) to weight the fit as:
\begin{equation}
\sigma_k=\sqrt{I_{{obs},k} - B + \left(\frac{R}{G}\right)^2}
\label{eq:Uncertainty}
\end{equation}

where $B$ (in digital number, DN) is the ADC (analog to digital converter) offset level, $G$ is the gain (in $e^-$/DN) and $R$ (in $e^-$) is the read noise of the image sensor. For our image sensor, the measured read noise is $R = 5.67 \ e^-$ and the gain is $G = 3.153\ e^-/{\rm DN}$. As the fitting results are not sensitive to the small changes in these terms, we used the average gain and the average read noise of the image sensor for the stable system. The routine requires an initial set of sensitivity parameters (all set to 1.0) which is then modified until a good fit is achieved.

%----------------
\section{Results}
\begin{figure}[ht]  
\centering \includegraphics[width=0.75\linewidth]{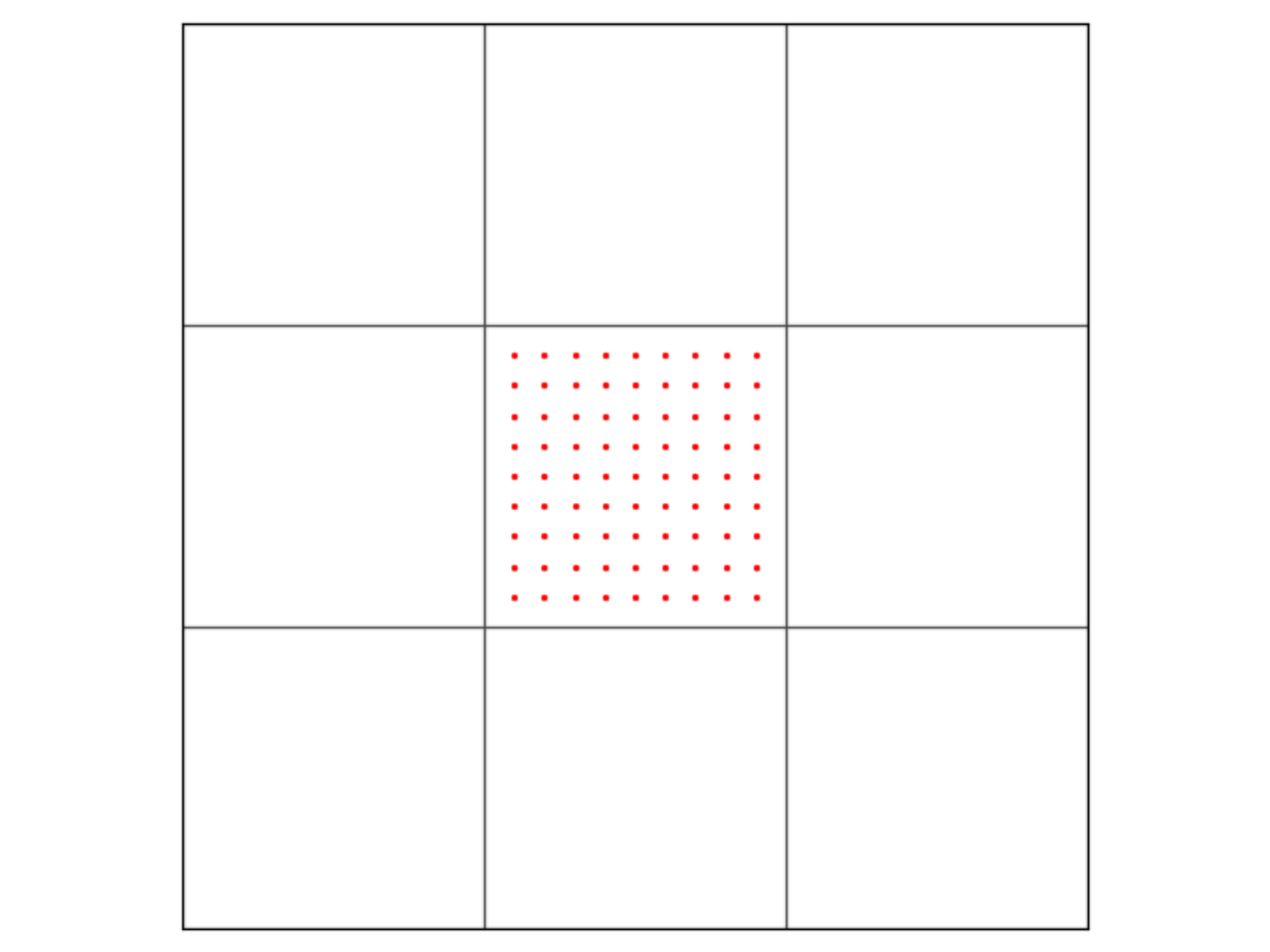}
\caption{\label{fig:IPSV_geo} Geometry of the selected scan region for IPS map extraction. Each square represents one pixel. The red dots ($9\times9$) represent the scanning region selected inside the POI for IPSV extraction.}
\end{figure}

\begin{figure}[ht] 
\centering \includegraphics[width=0.9\linewidth]{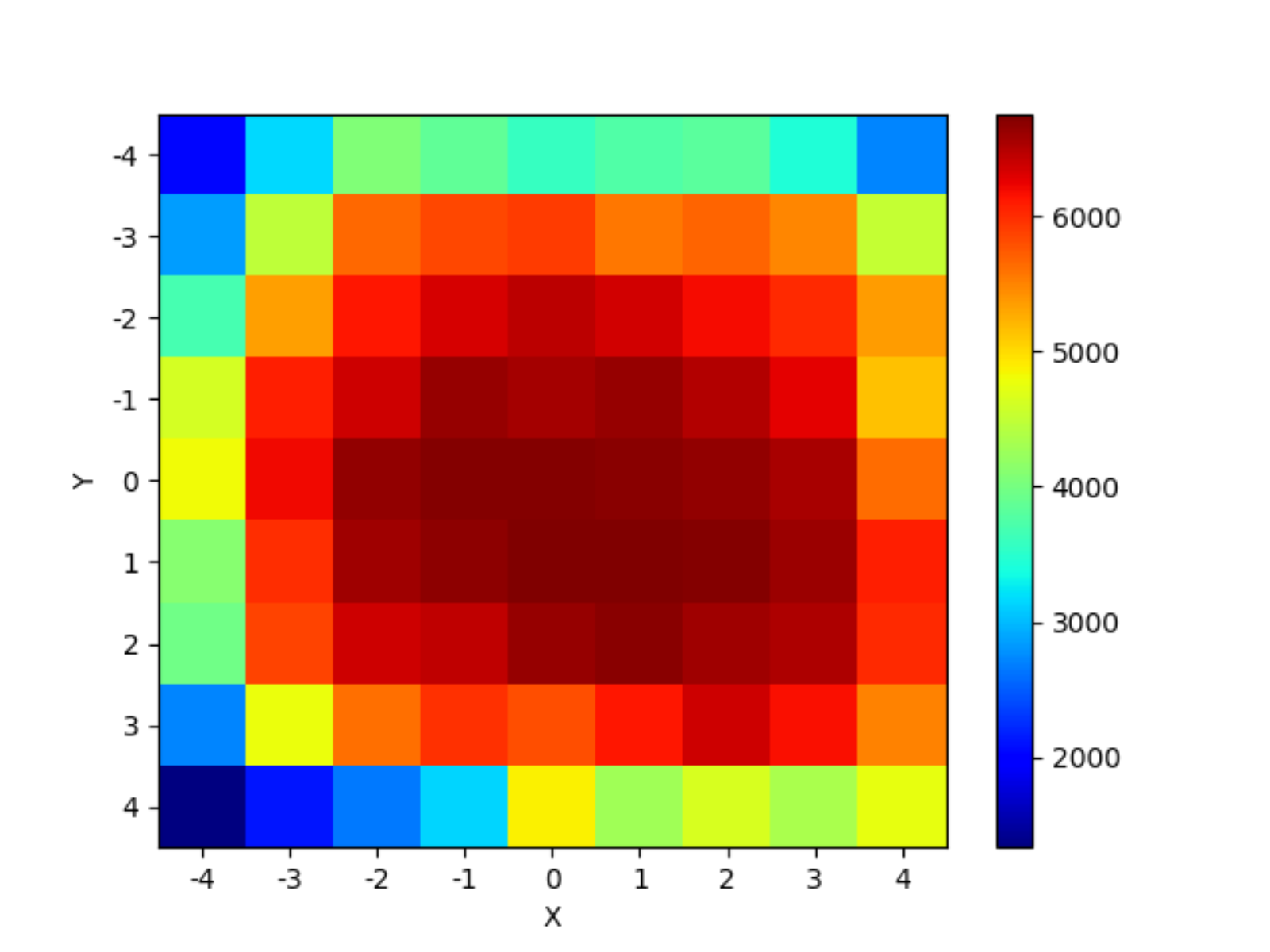}
\caption{The response of a single pixel (the POI) for each of the $(x,y)$ positions of the $9\times 9$ optical spots, aggregated into a 2D raster. These measurements were fitted to our model (\ref{eq:3}). The color scale represents the pixel signal value in digital number (DN).}
\label{fig:POI_response}
\end{figure}

%----------------
When the optical spot is completely inside the POI, the latter contains $\sim2\times10^4$ photoelectrons whereas neighboring pixels have only $10^2$ to $10^3$. At such high flux levels, the Poisson distribution of the observed POI image data tends to be Gaussian. We therefore choose all spot positions inside the POI so that the choice of a $\chi^2$ goodness-of-fit function (\ref{eq:Chi}) is justified. As shown in Fig.~\ref{fig:IPSV_geo}, we used $9\times9$ scan points inside the POI.  
As described in section II, the response of the POI for $9\times9$ scan positions is called the pixel response image as a function of scan positions, presented in Fig. \ref{fig:POI_response}. For the sake of clarity, Fig.~\ref{fig:POI_response} does not represent a $9\times 9$ pixel area of the sensor, but shows the responses of one single pixel, aggregated in a 2D raster.

\begin{figure}[ht] 
\centering \includegraphics[width=0.9\linewidth]{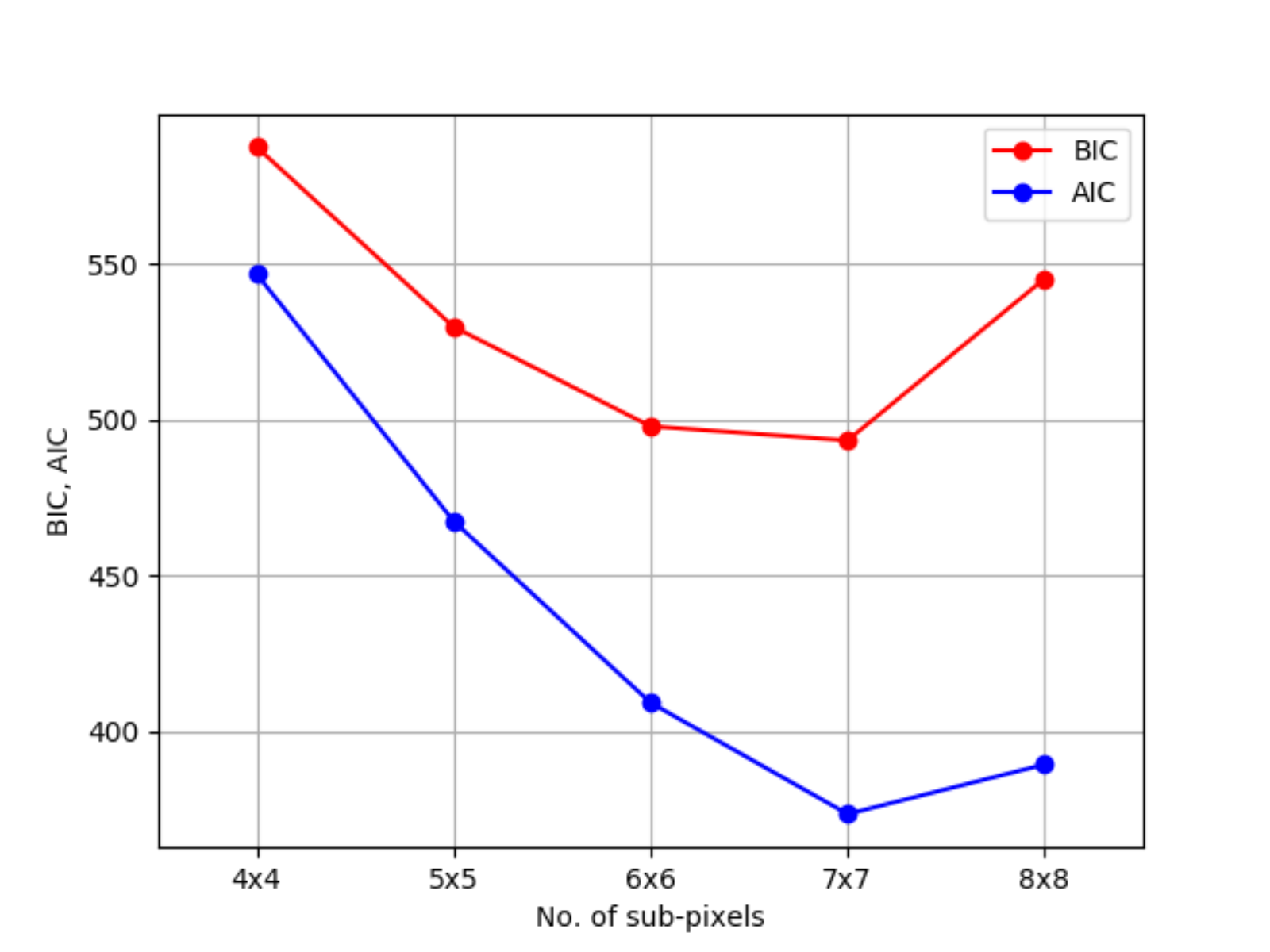}
\caption{\label{fig:bicAic} Optimal number of sub-pixel using the BIC and AIC over the POI image dataset for all scan positions.}
\end{figure}

\begin{figure}[ht] 
\centering \includegraphics[width=0.9\linewidth]{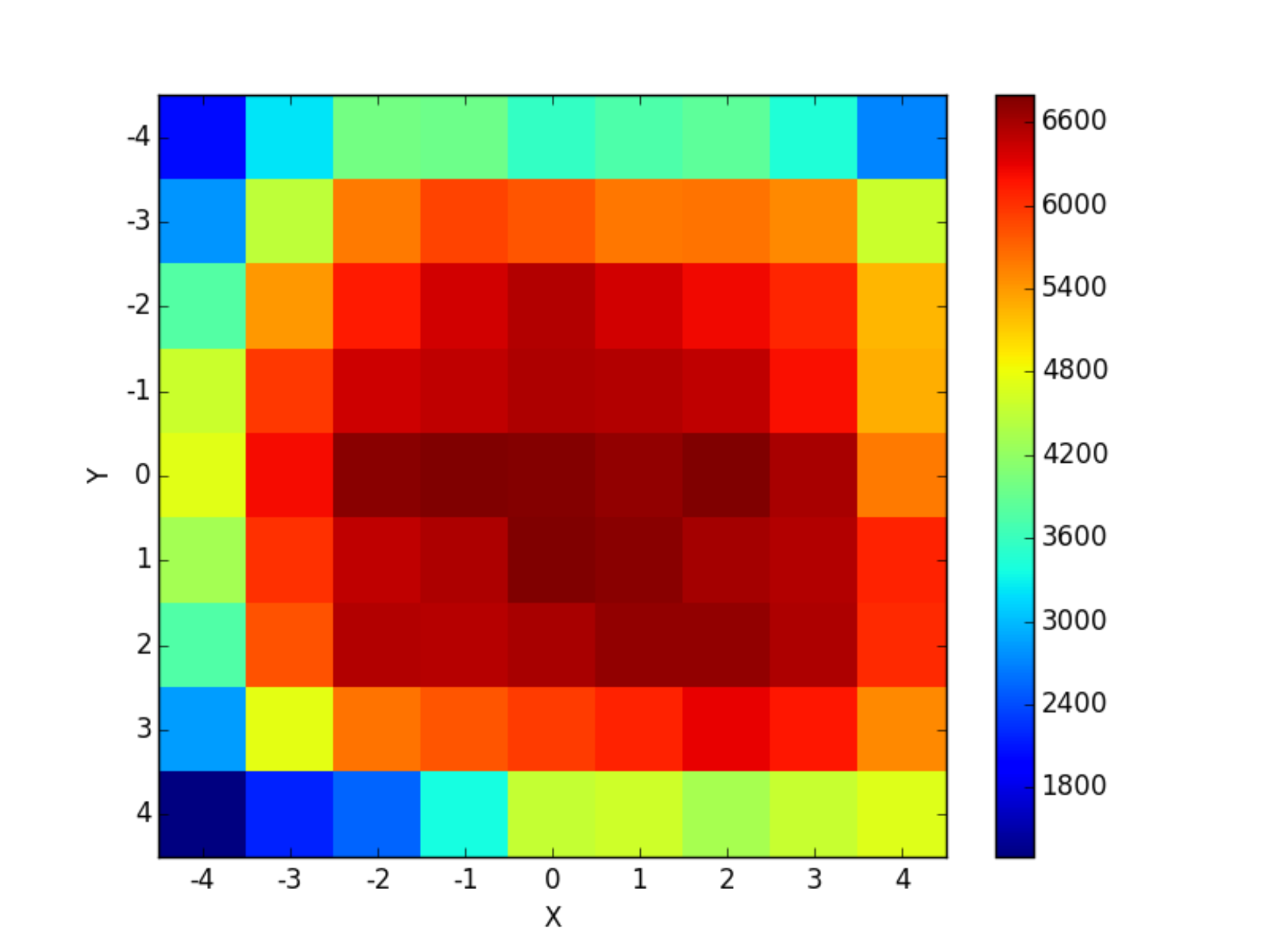}
\caption{\label{fig:ModelImage} The predicted pixel response values of the POI calculated from our best fit model, corresponding to the measurements in Fig.~(\ref{fig:POI_response}). The color scale represents the pixel signal value in digital number (DN).}
\end{figure}

To find the optimum number of sub-pixels to get an IPS map that is not overfitted nor underfitted, we used the Bayesian information criterion (BIC) and the Akaike information criterion (AIC). Both BIC and AIC address the problem of overfitting and underfitting by introducing a penalty term for the number of (sensitivity $S_{ij}$) parameters in the model \cite{IPSVref:R13}: 
\begin{eqnarray*}
  {\rm AIC} &=&  N \ln(\chi^2/N) + 2 p \\
  {\rm BIC} &=&  N \ln(\chi^2/N) + p \ln{N} \\
 \end{eqnarray*}
where $N$ is the number of data points (measured POI values for all scan positions), and $p$ is the number of variable parameters. 
We determined the optimal model and the corresponding BIC and AIC values for different choices for the number of $r\times r$ sub-pixels per pixel, and show the result in Fig.~\ref{fig:bicAic}. 
From this figure we deduce that for our dataset the optimal number of sub-pixel at which the BIC and the AIC curves attain a minimum is $7\times 7$. The reason why this number is \textit{smaller} than the $9\times 9$ spot positions is that the spots partly overlap. The information of the different POI response values is therefore not independent. Conclusively, we choose $7\times 7$ sub-pixels for our fitting to extract the IPS map. Our best-fitting model yields predicted pixel response values that are comparable to the measured ones, as shown in Fig.~ \ref{fig:ModelImage}.

\begin{figure}[ht] 
\centering \includegraphics[width=0.9\linewidth]{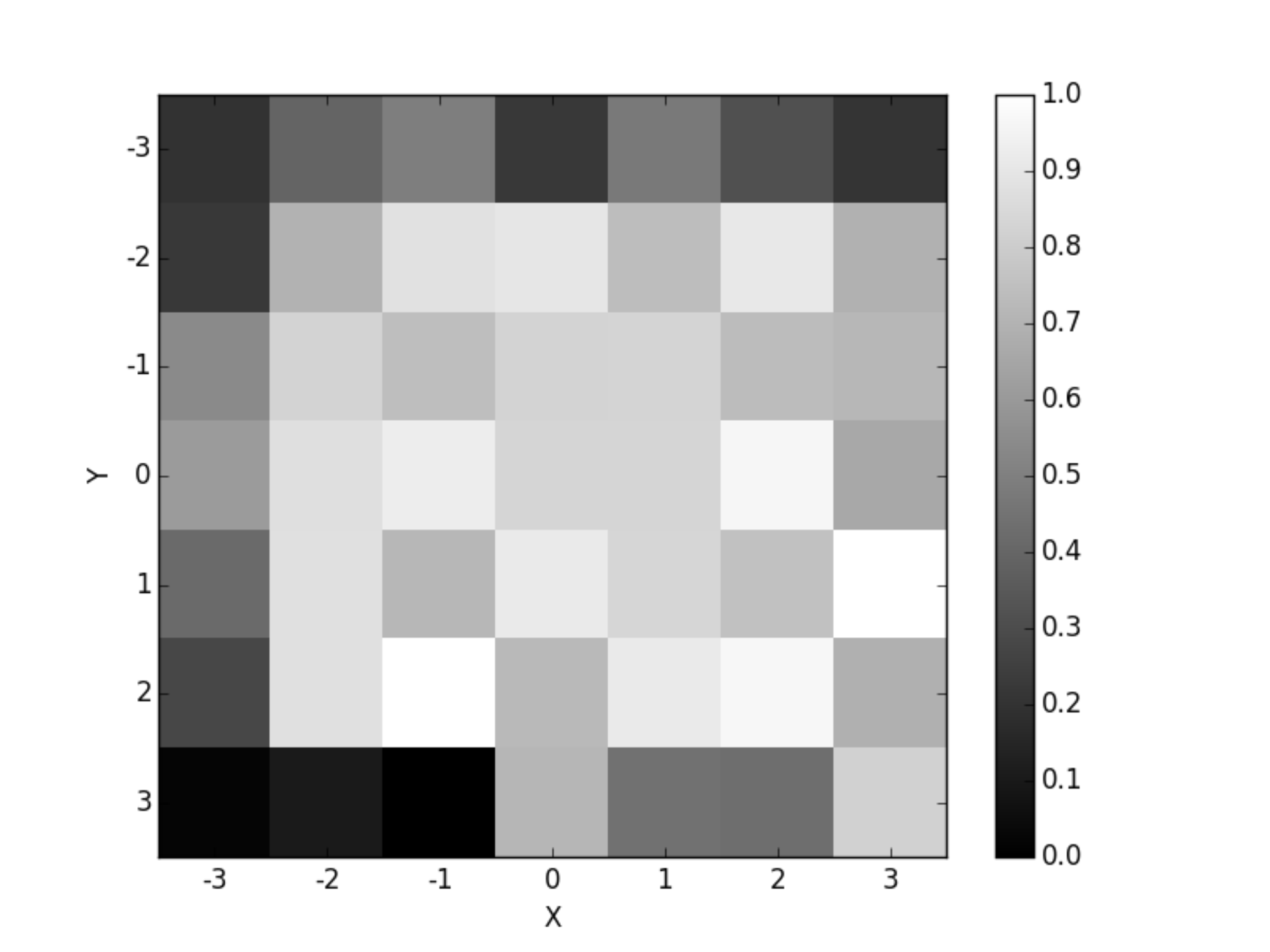}
\caption{\label{fig:IPSV_2D}The sensitivity for each of the $7\times 7$ sub-pixels of the POI. }
\end{figure}

The optimal sensitivity parameters $S_{ij}$ of our best model gives the most likely IPS map, and is presented in Fig.~\ref{fig:IPSV_2D}.
\begin{figure}[ht]
  \centering
  \subfigure[]{\includegraphics[width=0.9\linewidth]{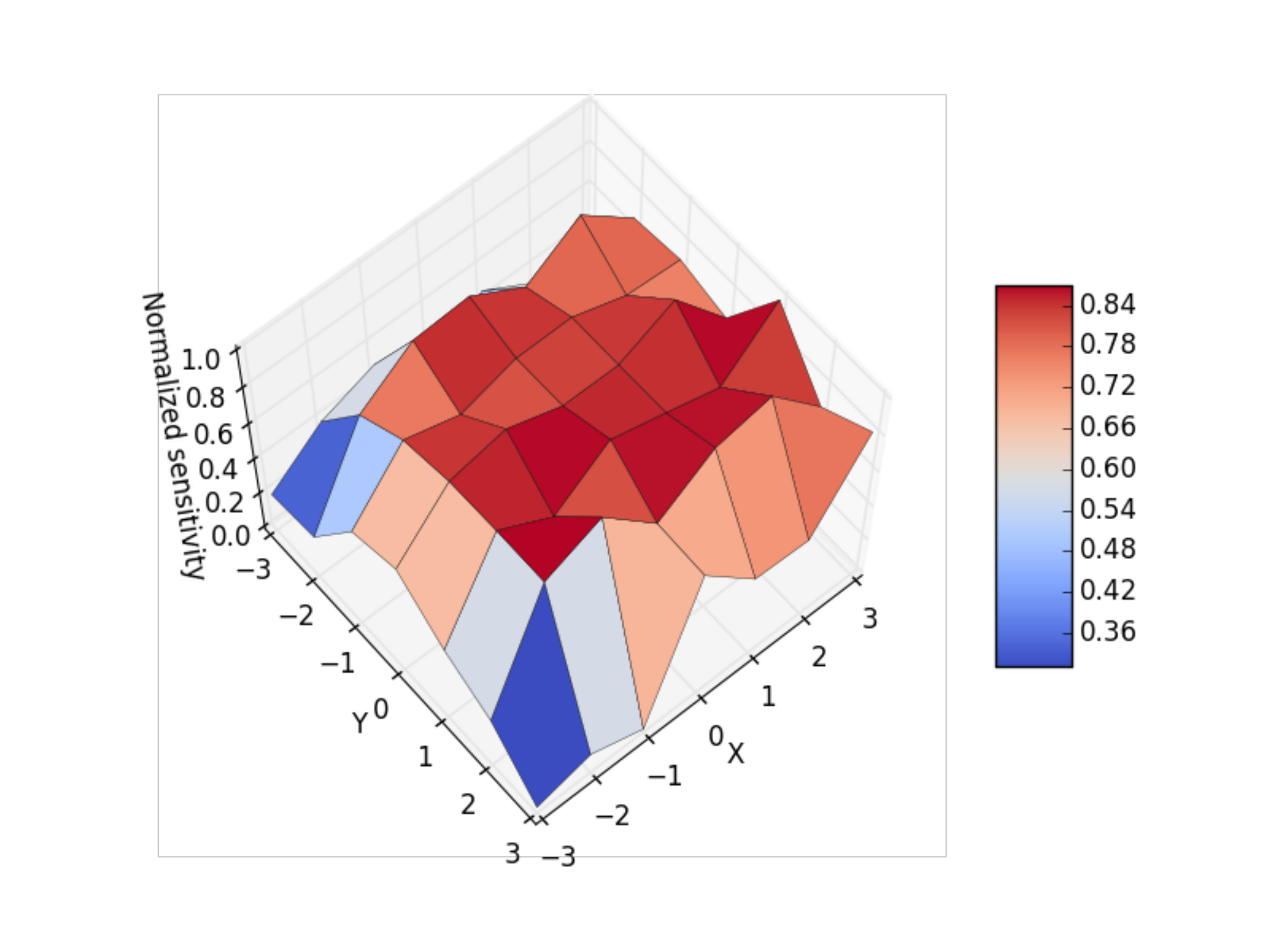}}
  \subfigure[]{\includegraphics[width=0.9\linewidth]{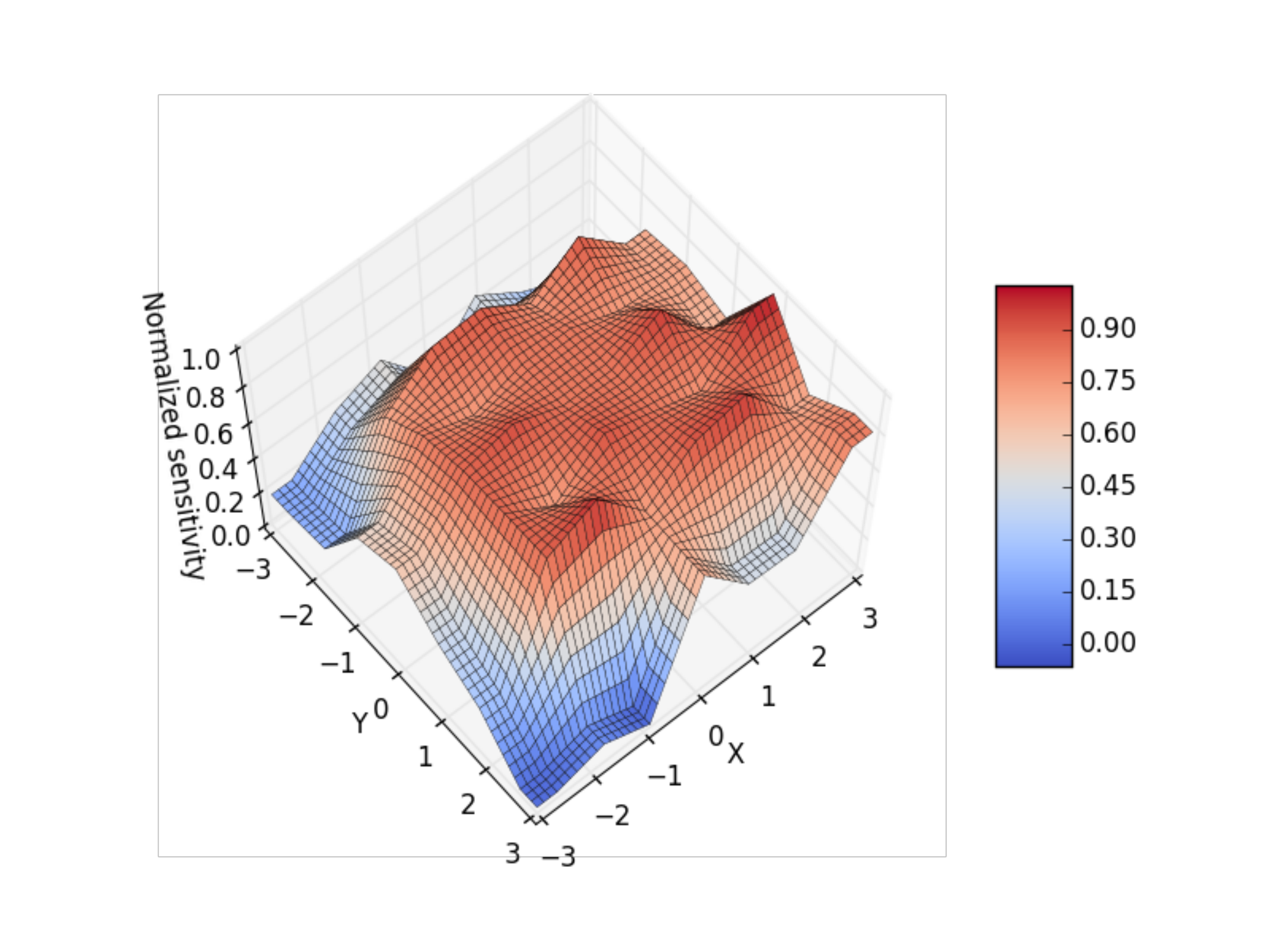}}
  \caption{\label{fig:IPSV_3D}(a) Surface plot of the intra-pixel sensitivity map of the POI. (b) Surface plot of the interpolated intra-pixel sensitivity map of the POI to better visualize the sensitivity variations.}
\end{figure}
The IPS map of the POI is plotted again as a surface plot in Fig. \ref{fig:IPSV_3D}(a) and an interpolated IPS map is presented in Fig. \ref{fig:IPSV_3D}(b), which provides a smoother transition of the sensitivity over the pixel area.  Clearly visible are the dips in the sensitivity at the edge of the pixel, especially at three adjacent edges. This is likely because the CMOS pixel does not have a $100\%$ fill-factor. The pixel grid includes the readout circuits at the edges and more or less symmetric around the sensitive area which reduces the sensitivity of the sub-pixels near the edges. We suspect that the sensitivity dip at the right-hand side edge is not visible due to a small horizontal misalignment of the spot positions, having an offset towards the left-hand side. We also apply the same method to retrieve the IPS map of some of the other pixels in the scan area, depicted in Fig. \ref{fig:ScanGeo}. Results of an another POI are given in the Appendix. Sensitivity dips along the edges are also visible there. 

Besides the sensitivity drop at the edges, there are also clear variations visible in the center of the pixel. To test whether these variations are genuine or whether they are merely noise, we formulate the following hypothesis test. The null hypothesis $H_0$ corresponds to an IPS map where each sub-pixel at the three adjacent edges (top, left, bottom) has its own sensitivity parameter $S_{ij}$, while all other sub-pixels have the same constant sensitivity $S_0$. The null model has therefore 21 free parameters: 19 $S_{ij}$ parameters for the sub-pixels at the 3 edges, 1 $S_0$ parameter for all other sub-pixels, and 1 $I_0$ parameter for the amplitude of the Airy disk function. The alternative hypothesis $H_1$ corresponds to a full model (Eq.~\ref{eq:3}), which has 49 sensitivity  parameters and one amplitude fit parameter. Next, we determine whether the full model gives a significantly better fit to the measured pixel response data than the $H_0$ model, using an F-test \cite{IPSVref:R22}:
\begin{equation}
F=\left(\frac{\chi^2_0-\chi^2_1}{\chi^2_1}\right) \cdot \left(\frac{N-p_1}{p_1-p_0}\right)
\end{equation}
where, N is the number of data points, $p_{0}=21$ is the number of parameters of the null hypothesis $H_0$ and $p_{1}=50$ is the number of parameters of the full hypothesis $H_1$. $\chi^2_0$ and $\chi^2_1$ are the weighted sums of squared residuals, corresponding to $H_0$ and $H_1$ respectively. We reject the null hypthesis if the P-value of the F-distribution is less than 0.05. In our case, the calculated F statistic from the data is 2.490 and the P-value is 0.007. This small P-value indicates that the null hypothesis does not hold and that the sensitivity at the pixel center likely cannot be assumed constant. Note, however, that the measured pixel response variations inside the pixel active area may not necessarily be silicon related, but could be due to microlens imperfections.  

\begin{figure}[ht] 
\centering \includegraphics[width=0.9\linewidth]{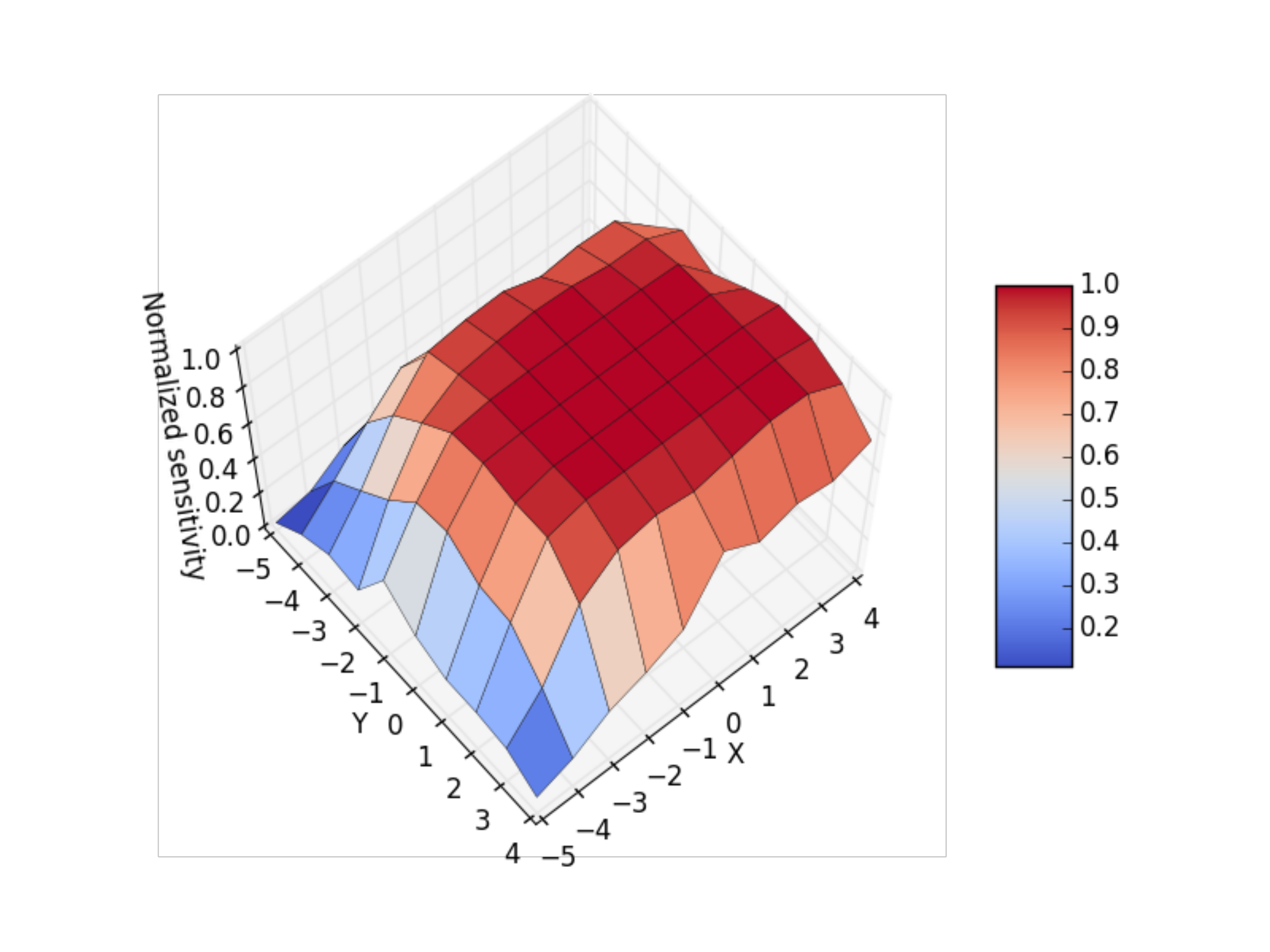}
\caption{\label{fig:IPSV_3D_Deconv}Surface plot of the intra-pixel sensitivity ($10\times10$ sub-pixel) of the POI, extracted by Wiener deconvolution method.}
\end{figure}
In many of the prior studies (\cite{IPSVref:R4}, \cite{IPSVref:R5}, \cite{IPSVref:R6}, \cite{IPSVref:R7}), the IPS variation is extracted by the deconvolution method. We therefore compare our forward modeling method with the deconvolution method. The latter exploits the fact that the measured pixel intensity profile is the convolution of instrumental PSF of our scanning system with the sub-pixel response function. One difficulty with direct deconvolution of the discretely-sampled pixel responses is that high-frequency noise tends to be amplified. One common method used to tackle this problem is Wiener deconvolution, which attempts to reduce the noise in a digital signal by suppressing frequencies with low signal-to-noise ratio. We applied Wiener deconvolution to remove the instrumental PSF from the pixel response data and show the result in Fig.~\ref{fig:IPSV_3D_Deconv}. The Wiener deconvolution seems unable to reconstruct the low-amplitude high-frequency components, but rather suppresses them. As a result, information on small sensitivity variations in the center of pixel is lost.

%-------------------------------
\section{Summary and Conclusion}
%-------------------------------
In this paper we measured the IPS map of a CMOS image sensor. We presented a forward modeling method to extract the sensitivity variations over a single pixel area. We modeled the optical spot projection system, discretized the sub-pixel response function, and used least-squares fitting to model to the measurements. One advantage of this forward technique is that it allows including the photon noise and the readout noise. The results show that the sensitivity clearly varies inside a pixel, with dips at the edges of the pixel. Moreover, we show with our forward modeling approach that small intra-pixel sensitivity variations in the center of the pixel are also significant. To test whether these variations are genuine or whether they are merely noise, we carried out a hypothesis test and validated the IPS variation in the center of the pixel. Finally, we compare our results to those obtained from the more commonly used Wiener deconvolution and conclude that this method is unable to resolve these small central sensitivity variations.

The resulting IPS was obtained for several pixels. Providing the IPS is obtained for all relevant pixels in the detector, our results can be used to calibrate temporal noise in time series of photometric fluxes, originating from a PSF jittering over a pixel due to an imperfect attitude control system.

%--------
\appendix
%--------
\begin{figure}[!htp] 
\centering \includegraphics[width=0.9\linewidth]{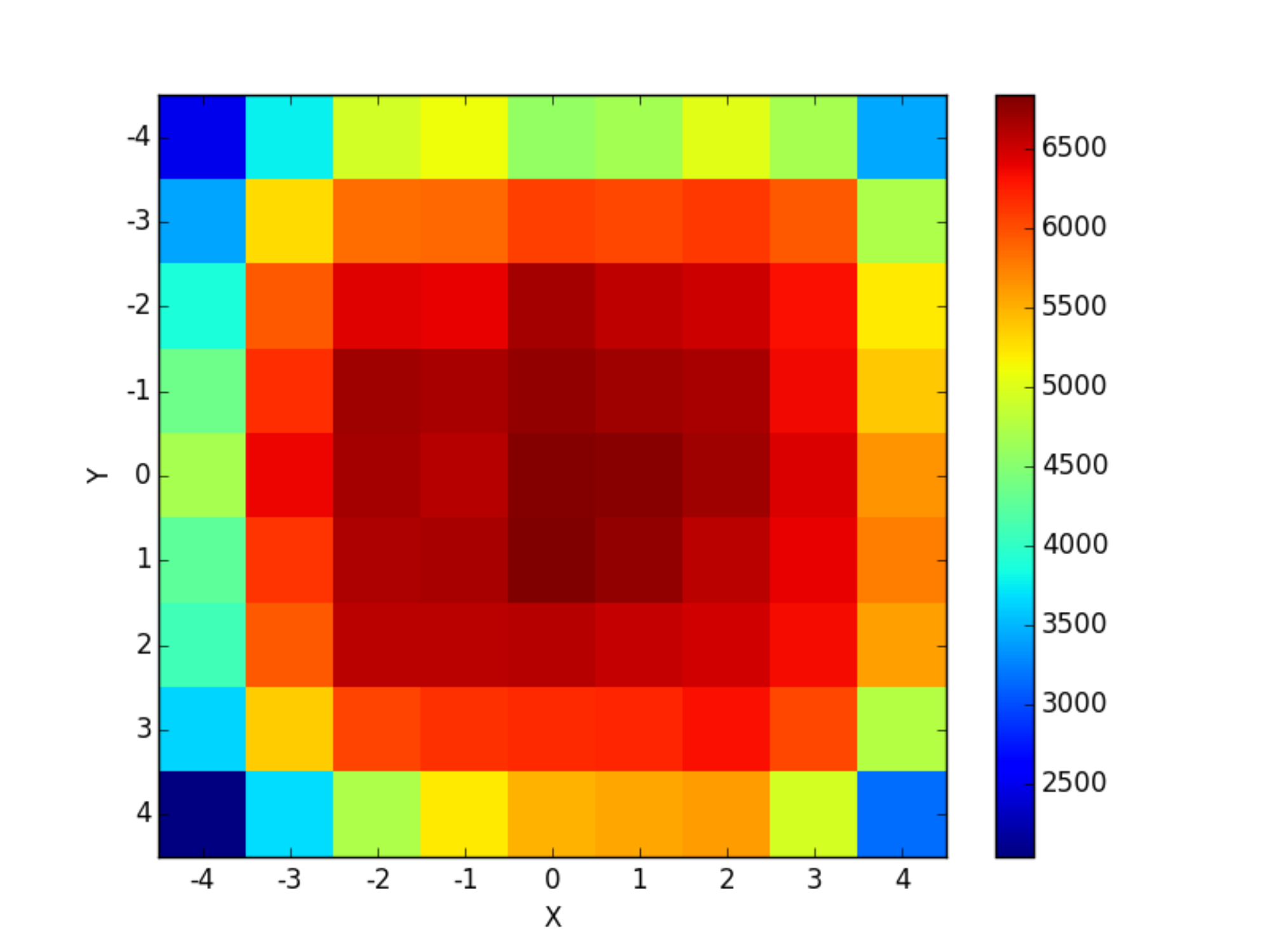}
\caption{\label{fig:ModelImage_3_3} The predicted pixel response values of another POI calculated from our best fit model. The color scale represents the pixel signal value in digital number (DN).}
\end{figure}

\begin{figure}[!htp] 
\centering \includegraphics[width=0.9\linewidth]{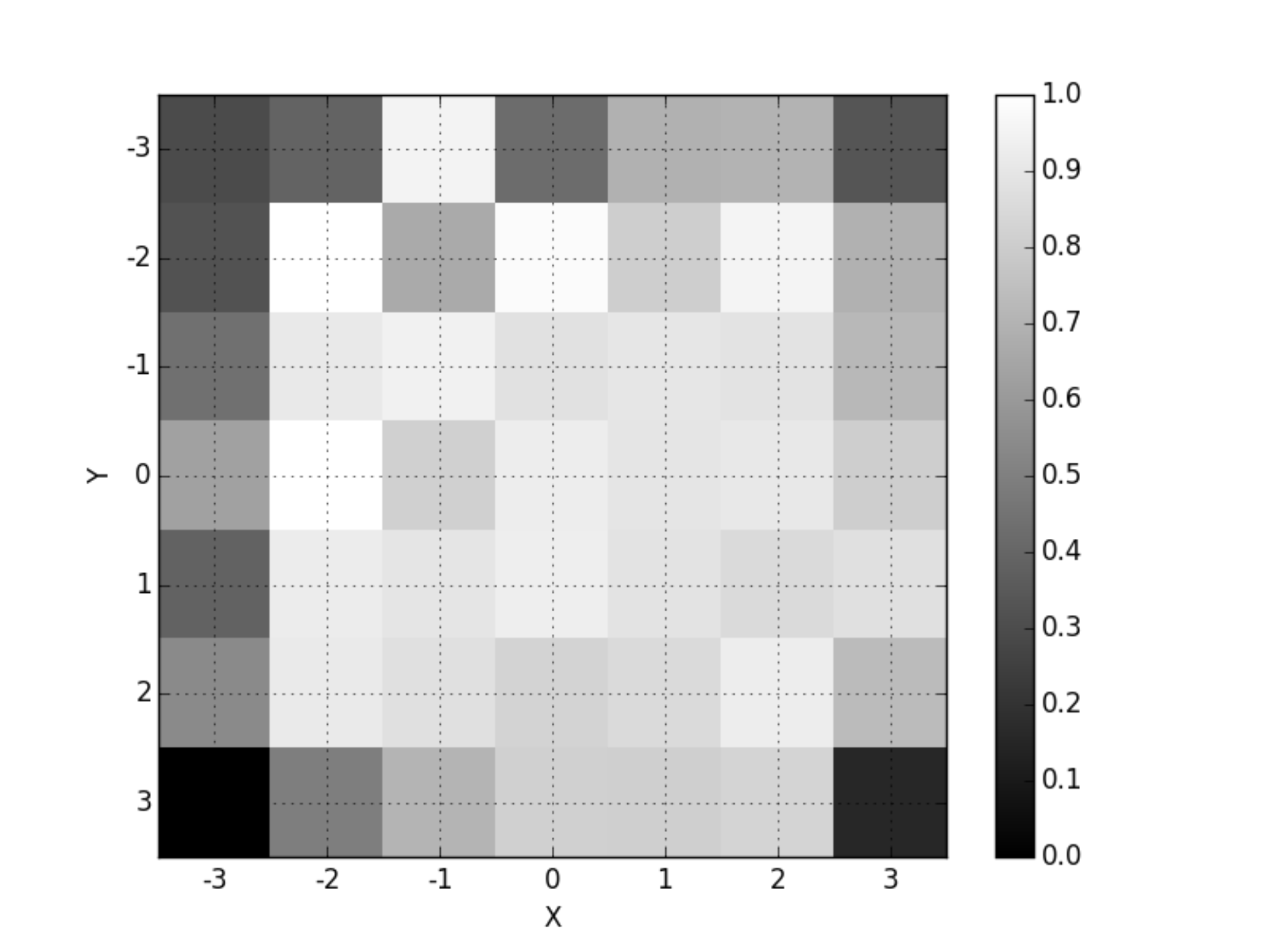}
\caption{\label{fig:IPSV_2D_3_3}The sensitivity for each of the $7\times 7$ sub-pixels of another POI.}
\end{figure}

%------------------------
\section*{Acknowledgment}
%------------------------

This work is funded by IWT-Baekeland. The authors would like to thank CMOSIS (now ams Sensors, Belgium) for the measurement setup and their technical support.

% Can use something like this to put references on a page
% by themselves when using endfloat and the captionsoff option.
\ifCLASSOPTIONcaptionsoff
  \newpage
\fi

\bibliographystyle{IEEEtran}
\bibliography{IPSVref}

% biography section
% 
% If you have an EPS/PDF photo (graphicx package needed) extra braces are
% needed around the contents of the optional argument to biography to prevent
% the LaTeX parser from getting confused when it sees the complicated
% \includegraphics command within an optional argument. (You could create
% your own custom macro containing the \includegraphics command to make things
% simpler here.)
%\begin{IEEEbiography}[{\includegraphics[width=1in,height=1.25in,clip,keepaspectratio]{mshell}}]{Michael Shell}
% or if you just want to reserve a space for a photo:

% You can push biographies down or up by placing
% a \vfill before or after them. The appropriate
% use of \vfill depends on what kind of text is
% on the last page and whether or not the columns
% are being equalized.

%\vfill

% Can be used to pull up biographies so that the bottom of the last one
% is flush with the other column.
%\enlargethispage{-5in}

% that's all folks
\end{document}